# The Physical Properties of ThCr$_2$Si$_2$-type Nickel-based Superconductors BaNi$_2$T$_2$ (T = P, As): An *ab-initio* study

**Md. Atikur Rahman[1]\*, Md. Zahidur Rahaman[2], Md. Lokman Ali[1], Md. Shahjahan Ali[1]**

[1]Department of Physics, Pabna University of Science and Technology, Pabna, 6600, Bangladesh
[2]Department of Physics, Bangladesh University of Engineering and Technology, Dhaka, Bangladesh

**Abstract**

Employing the first-principles computations based on the density functional theory (DFT), we have investigated the structural, mechanical, electronic, bonding, optical and thermodynamics properties of newly discovered bulk superconductors BaNi$_2$P$_2$ ($T_c$ ~3 K) and BaNi$_2$As$_2$ ($T_c$ ~0.7 K). Our optimized lattice parameters are in good concurrence with the experimental records. The positive elastic constants reveal that both the superconductors are stable in nature. The analysis of the mechanical properties insures that both the phases are ductile in nature and show anisotropic behaviors. The analysis of the electronic band structures and density of states (TDOS and PDOS) reveals the metallic manners for both the compounds and the major contribution comes from Ni-3d states for both the phases. The calculations of the chemical bonding specify that a mixture of covalent, ionic and metallic bonds exist in both the superconductors. The high value of the dielectric constant of BaNi$_2$As$_2$ ensures that this compound may be used to manufacture high value capacitors. The large reflectivity in the low energy region indicates that both the compounds might be useful as coating materials to reduce solar heating. By using the elastic constants data the calculated Debye temperatures of BaNi$_2$P$_2$ and BaNi$_2$As$_2$ are 323.70 K and 272.94 K respectively, which are in good accordance with the experimental values. Finally we have calculated the thermal conductivity of these compounds which is 0.56 for BaNi$_2$P$_2$ and 0.46 for BaNi$_2$As$_2$.

\*Corresponding author: atik0707phy@gmail.com

# 1. Introduction

$ThCr_2Si_2$ type ternary intermetallic materials usually accommodate a superconducting ground state. The rare-earth $AM_2X_2$ structural materials have received great interest of researchers because of their many rich characteristics. This type of compounds usually possesses superconductivity, exotic magnetic order, valence fluctuation and heavy fermion nature [1]. The presences of strongly correlated *d*-electrons in the transition metal compounds are usually responsible for many intriguing physical characteristics including superconductivity of this type of compounds [2-4]. Even in the family of Fe-based superconductors, the parent compounds $AFe_2As_2$ (A = Sr, Ca, Ba etc.) possesses the $ThCr_2Si_2$ type structure [5]. At low temperature they belong to SDW (Spin Density Wave) semimetals. For solving the hidden pairing mechanism in the high temperature superconductors many current research works are focused on the $ThCr_2Si_2$ type ternary intermetallic materials because a high quality sample of single crystal can easily be synthesized for these intermetallics [6-9]. Moreover, $ThCr_2Si_2$ type simple tetragonal body-centered structure can be formed by most of the transition metal. To investigate the superconducting nature among the similar compounds within the $ThCr_2Si_2$ type structure may be a simple way for exploring new superconductors.

In the family of Fe-based superconductor, 122-type ternary intermetallics $AFe_2As_2$ (A = Ca, Sr, Ba etc.) possess metallic character and are usually free of oxygen [10-12]. For exploring the superconducting mechanism in these materials, many current research works are focused on these $ThCr_2Si_2$ type ternary intermetallic materials [13, 14]. The transition metal, nickel (Ni) is located in the same period of iron (Fe) and so we can expect similar superconductivity in Ni-based compounds. Among the nickel based superconductors, $BaNi_2P_2$ was first synthesized in 1997 by Keimes *et al* through heating element mixture and characterized by X-ray diffraction method [15]. Later in 2008 Mine *et al* reported the infinite layer structure of $BaNi_2P_2$ with superconducting critical temperature of 3 K [16]. They found bulk superconductivity in $BaNi_2P_2$. In 4 to 300 K $BaNi_2P_2$ exhibits Pauli paramagnetism and metallic conduction [16]. In 2009 Shein *et al* studied the electronic and structural properties of $ANi_2X_2$ (A = Ba, Sr and X = P, As) by using first-principles method [17]. They found complex anisotropic bonding character in $ANi_2X_2$ (A = Ba, Sr and X = P, As). $BaNi_2As_2$ intermetallic was synthesized by Ronning *et al* in 2008 with $ThCr_2Si_2$ type structure [18]. This compound shows thermal hysteresis of 7 K and first order phase transition at 130 K [18]. Bulk superconductivity is observed in $BaNi_2As_2$ with superconducting transition temperature 0.7 K [18, 20]. In 2009 Sefat *et al* also reported the structural phase transition of $BaNi_2As_2$ at 131 K with superconductivity at 0.69 K [19]. $BaNi_2As_2$ is a conventional phonon-mediated superconductor with a large Fermi surface [21].

Though remarkable progress has been made to investigate the electronic properties and superconducting mechanism of $BaNi_2P_2$ and $BaNi_2As_2$ ternary intermetallic compounds, there is still lack of investigation about the detailed physical properties of these two superconductors. In our previous work, we have carried out a thorough investigation about the detailed physical properties of $LaRu_2M_2$ (M = P and As) and $BaIr_2Mi_2$ (Mi = P and As) superconductors by theoretical method [22, 23]. In this study, we aim to investigate the detailed physical properties including structural, electronic, optical, elastic and thermodynamic properties of ternary intermetallic superconductors $BaNi_2P_2$ and $BaNi_2As_2$ by using Density Functional Theory (DFT) based theoretical method. A thorough comparison with a proper discussion has also been done among the evaluated properties of these two superconductors.

## 2. Computational aspects

All of our theoretical investigations based on the DFT calculations [24] were executed in CASTEP code [25] utilizing the generalized gradient estimation to the exchange-correlation potential with Perdew-Burke-Ernzerhof (PBE) system [26]. For pseudo atomic calculations, the $P-3s^2\ 3p^3$, $Ni-3d^8\ 4s^2$ and $Ba-5s^2\ 5p^6\ 6s^2$ for $BaNi_2P_2$ and $As-4s^2\ 4p^3$, $Ni-3d^8\ 4s^2$ and $Ba-5s^2\ 5p^6\ 6s^2$ in case of $BaNi_2As_2$ have been taken as valance electrons. The plane wave energy cutoff of 550 eV for $BaNi_2P_2$ and 500 eV for $BaNi_2As_2$ were used to extend the wave functions. In order to Brillouin zone sampling a Monkhorst-Pack grid [27] of 10×10×12 k-points are used for $BaNi_2P_2$ and $BaNi_2As_2$ compounds. Using Broyden-Fletcher-Goldfarb-Shanno (BFGS) minimization manner [28], we have optimized the geometric structures of both superconductors. For geometry optimization the succeeding criteria were placed to $1 \times 10^{-5}$ eV/atom for total energy, 0.003 eV/Å for maximum force, 0.005 GPa for maximum stress and 0.001 Å for maximum atomic displacement. To calculate approximately the single and polycrystalline elastic constants of these compounds, we have used the Voigt-Reuss-Hill approximation [29]. For convenience 12×12×12 k-points have been used to calculate the elastic constants of $BaNi_2As_2$ superconductor.

## 3. Results and Discussion

### 3.1 Structural properties

At room temperature, the ternary compounds $BaNi_2T_2$ (T = P, As) have $ThCr_2Si_2$-type layered tetragonal crystal structures with the space group of I4/mmm (no.139) [30, 31]. The conventional and the optimized crystal structures of both these superconductors [type-II] are represented in Fig. 1. Each unit cell have two formula units (Z = 2) with ten atoms that means one formula unit for each primitive cell with five atoms. The Wyckoff atomic positions of both the phases are: Ba: 2a (0, 0, 0), Ni: 4d (0, ½, ¼) and (P, As): 4e (0, 0, $Z_A$) where $Z_A$ is known as internal co-ordinate [32-34]. $Z_A$ = 0.3431 for $BaNi_2P_2$ and $Z_A$ = 0.3476 for the compound BaNi$_2$As$_2$. The optimized lattice parameters at zero pressure $a_0$, $c_0$; the equilibrium cell volume $V_0$ and the bulk modulus $B_0$ with the available experimental data are listed in Table 1.

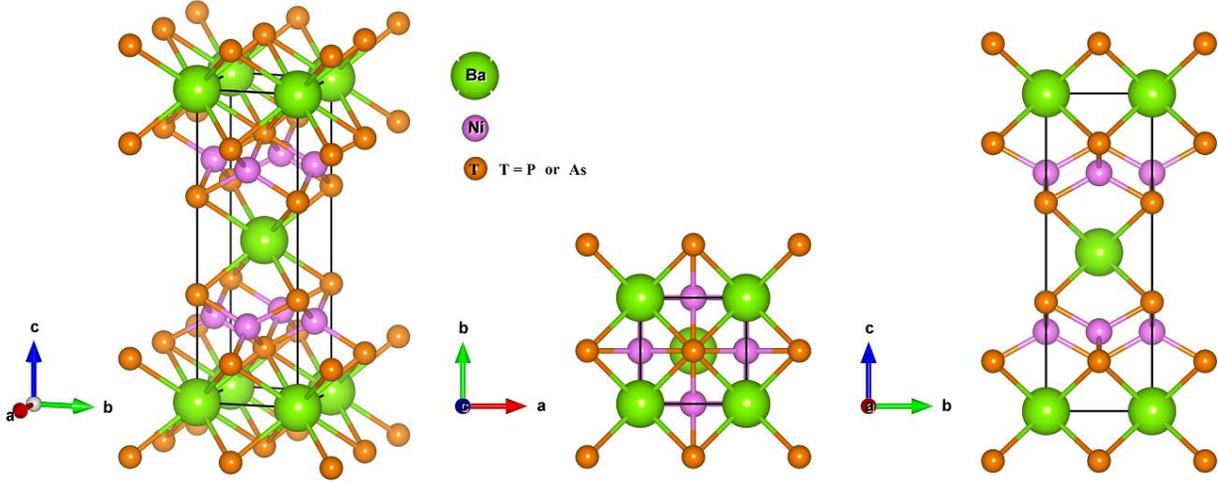

**Fig. 1.** Three and two dimensional crystal structures of BaNi$_2$T$_2$ (T = P or As).

**Table 1:** The observed equilibrium lattice parameters, unit cell volume, and bulk modulus of ThCr$_2$Si$_2$-type BaNi$_2$P$_2$ and BaNi$_2$As$_2$ superconductors.

| Properties | BaNi$_2$P$_2$ | | Deviation from Expt. (%) | BaNi$_2$As$_2$ | | Deviation from Expt. (%) |
|---|---|---|---|---|---|---|
| | This work | Expt.[32] | | This work | Expt.[33,34] | |
| $a_0$ (Å) | 3.939 | 3.947 | 0.203 | 4.149 | 4.112 | 0.892 |
| $c_0$ (Å) | 11.957 | 11.820 | 1.146 | 11.834 | 11.540 | 2.484 |
| $c_0/a_0$ | 3.04 | 2.99 | 1.672 | 2.85 | 2.81 | 1.423 |
| $V_0$ (Å$^3$) | 185.52 | - | - | 203.71 | - | - |
| $B_0$ (Gpa) | 111.67 | - | - | 121.98 | - | - |

Analyzing Table 1 it is clear that the calculated lattice parameters are approximately close to the experimental values. This makes sure the reliability of the DFT-based *ab-initio* investigations. From Table 1 we have seen that, our studied lattice parameters are slightly deviated from the experimental results. This is due to the temperature dependency of the lattice constants and GGA route [35].

**3.2 Mechanical Properties**

The elastic constant is a fundamental parameter which gives key information about the nature of force in a crystal. It is related to the various solid state phenomena such as ductility, brittleness, mechanical stability, anisotropy and stiffness of solids [36]. In this section, we have studied the mechanical behavior of BaNi$_2$P$_2$ and BaNi$_2$As$_2$ superconductors. The elastic constants of both the semiconductors were calculated using the static finite strain method [37]. A tetragonal crystal system has six independent elastic constants namely; $C_{11}$, $C_{12}$, $C_{13}$, $C_{33}$, $C_{44}$ and $C_{66}$. In Table 2, we have presented the calculated values of the six individual elastic constants. For tetragonal crystal structure, the elastic constants need to satisfy the mechanical stability conditions known as the Born stability criteria [38].

$$\left.\begin{array}{l} C_{11} > 0,\ C_{44} > 0,\ C_{33} > 0,\ C_{66} > 0 \\ C_{11} + C_{33} - 2C_{13} > 0,\ C_{11} - C_{12} > 0 \\ 2(C_{11} + C_{12}) + C_{33} + 4C_{13} > 0 \end{array}\right\} \quad (1)$$

The calculated elastic constants (Table 2) are all positive and satisfy the above generalized conditions indicating that BaNi$_2$P$_2$ and BaNi$_2$As$_2$ superconductors are mechanical stable in nature. To the best of our knowledge, there are no experimental and theoretical data available about the elastic constants for both the superconductors. We consider the present results as prediction study provide a useful reference for future work.

**Table 2.** The calculated elastic constants $C_{ij}$ (in GPa) of BaNi$_2$P$_2$ and BaNi$_2$As$_2$ superconductors.

| Compounds | Elastic constants | | | | | |
|---|---|---|---|---|---|---|
| | $C_{11}$ | $C_{12}$ | $C_{13}$ | $C_{33}$ | $C_{44}$ | $C_{66}$ |
| BaNi$_2$P$_2$ | 171.91 | 21.23 | 66.51 | 89.25 | 43.84 | 23.04 |
| BaNi$_2$As$_2$ | 162.47 | 34.89 | 66.23 | 93.72 | 42.91 | 17.63 |

Using the calculated elastic constants, the bulk modulus, shear modulus, Young's modulus, anisotropy, and hardness for the both superconductors are determined by the Voigt-Reuss-Hill approximations [39]. Based on the Voigt-Reuss-Hill approximations, the bulk and shear modulus for tetragonal structure are as follows:

$$B_V = \frac{2C_{11} + 2C_{12} + C_{33} + 4C_{13}}{9} \quad (2)$$

$$B_R = \frac{C^2}{M} \quad (3)$$

$$G_V = \frac{M + 3C_{11} - 3C_{12} + 12C_{44} + 6C_{66}}{30} \quad (4)$$

$$G_R = \frac{15}{\left[\frac{18B_V}{C^2} + \frac{6}{(C_{11} - C_{12})} + \frac{6}{C_{44}} + \frac{3}{C_{66}}\right]} \tag{5}$$

Where,

$$M = C_{11} + C_{12} + 2C_{33} - 4C_{13}$$

And

$$C^2 = (C_{11} + C_{12})C_{33} - 2C_{13}^2$$

Hill took an average value of $B$ and $G$ given by the following two relations,

$$B = \frac{1}{2}(B_R + B_v) \tag{6}$$

$$G = \frac{1}{2}(G_v + G_R) \tag{7}$$

Now we can calculate the Young's modulus ($E$), and Poisson's ratio ($v$) from the below relations,

$$E = \frac{9GB}{3B + G} \tag{8}$$

$$v = \frac{3B - 2G}{2(3B + G)} \tag{9}$$

The anisotropic behaviors of a crystal can be estimated by the following equation [40],

$$A^U = \frac{5G_V}{G_R} + \frac{B_V}{B_R} - 6 \tag{10}$$

The hardness of materials can be predicted by using the following relation [41],

$$H_V = 2(K^2 G)^{0.585} - 3 \tag{11}$$

The calculated polycrystalline elastic parameters by using Eq. 2 to Eq. 11 are tabulated in Table 3. According to Table 3, the evaluated values of bulk modulus of both compounds are lower than 100 GPa and hence, both these compounds should be classified as a relatively soft materials. The shear modulus is another interesting property which represents the resistance to shear deformation against external forces. The calculated values of shear modulus for $BaNi_2P_2$ are higher than that of $BaNi_2As_2$ superconductor. It is well known that, the large value of shear modulus is a parameter of the high directional bonding between atoms [42]. The present calculated values indicating that, the directional bonding between the atoms of $BaNi_2P_2$ is quite small. The Young's modulus is defined as the ratio of the tensile stress to tensile strain, which measure the stiffness for solid material. The higher value makes the solid stiffer. In this calculation, we have seen that the value of Young's modulus of $BaNi_2P_2$ is larger than the $BaNi_2As_2$ indicating that the compound $BaNi_2P_2$ is stiffer than $BaNi_2As_2$.

According to Pugh criterion [43], the B/G ratio is used to separate the ductility or brittleness manner of a material. If B/G is greater than 1.75, the materials behave in a ductile nature and if the value is less than 1.75, the material behaves in brittle manner. From Table 3 we see that the B/G ratio of both the superconductors is greater than 1.75, so both the compounds behave as ductile manner. The Poisson's ratio is used to measure the stability of the materials against shear and provides important information about the nature of the bonding force [44]. P. *Ravindran et. al.* [42] reported that the interval of Poisson's ratio for central force materials is 0.25-0.50. In the present case, the obtained values of Poisson's ratio indicate the existence of central force in both the superconductors. The elastic anisotropy factor has an important implication in material science as well as in crystal physics. It provides a measure of the degree of anisotropy of the materials [40].

**Table 3.** Calculated polycrysttalline bulk modulus *B* (GPa), shear modulus *G* (GPa), Young's modulus *E* (GPa), *B/G* values, Poisson's ratio *v*, elastic anisotropy $A^U$ and Vickers hardness $H_v$ (GPa) of $BaNi_2P_2$ and $BaNi_2As_2$ superconductors.

| Polycrystalline elastic properties | | | | | | | |
|---|---|---|---|---|---|---|---|
| Compounds | *B* | *G* | *E* | *B/G* | *v* | $A^U$ | $H_v$ |
| $BaNi_2P_2$ | 80.92 | 35.87 | 93.77 | 2.25 | 0.30 | 1.60 | 3.27 |
| $BaNi_2As_2$ | 82.40 | 33.37 | 88.21 | 2.46 | 0.32 | 1.41 | 2.40 |

For complete isotropic materials, the value of $A^U$ is zero otherwise the material will be anisotropic. From Table 3 we see that both compounds show large anisotropic characteristics whereas $BaNi_2P_2$ shows large anisotropy than $BaNi_2As_2$. *Chen. et. al.* [41] gives a new theoretical model to estimate the hardness of polycrystalline crystals. According to Table 3, the obtained low values of $H_v$ reveal that both the compounds can be classified as soft materials.

## 3.3 Electronic Properties

Fig. 2 (a) and (b) shows the electronic band structures of $BaNi_2P_2$ and $BaNi_2As_2$ respectively toward the high symmetry direction in the Brillouin zone. The electronic band structure provides significant information about a material to be metal, semiconductor or insulator. The bonding features of a material are obtained from the partial and total density of states calculation [45]. It is evident from these band structures that several valence bands across the Fermi level and overlap with the conduction bands indicating no band gap at the Fermi level. This analysis reveals that both the phases under study show metallic feature. The metallic behaviors of $BaNi_2P_2$ and $BaNi_2As_2$ offer a hint that these compounds may be superconductor [46]. The partial and total density of states for the primitive cell of $BaNi_2T_2$ (T= P, As) are displayed in Fig. 3 (a) and 3 (b) respectively.

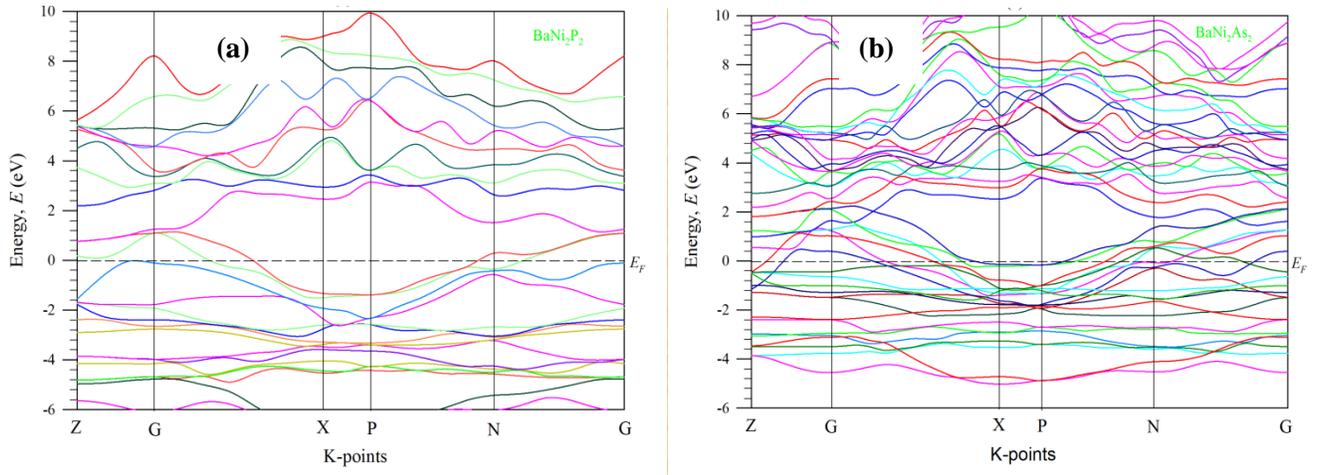

**Fig. 2.** The electronic band structures of (a) $BaNi_2P_2$ and (b) $BaNi_2As_2$ superconductors along high symmetry direction in the Brillouin zones.

The contributions of partial and total density of states at the Fermi level also indicating the metallic behavior of these two phases. In the band structures of $BaNi_2P_2$ and $BaNi_2As_2$, there is a lowest band lying in the energy region -10 eV below the Fermi level $E_F$, which arises mainly from P-3s states for $BaNi_2P_2$ and As-4s states for $BaNi_2As_2$ and are separated from the near-Fermi valence bands by a gap. These bands are to be found in the energy range from -6 eV to the Fermi level and formed mainly from P-3p and Ni-3d states.

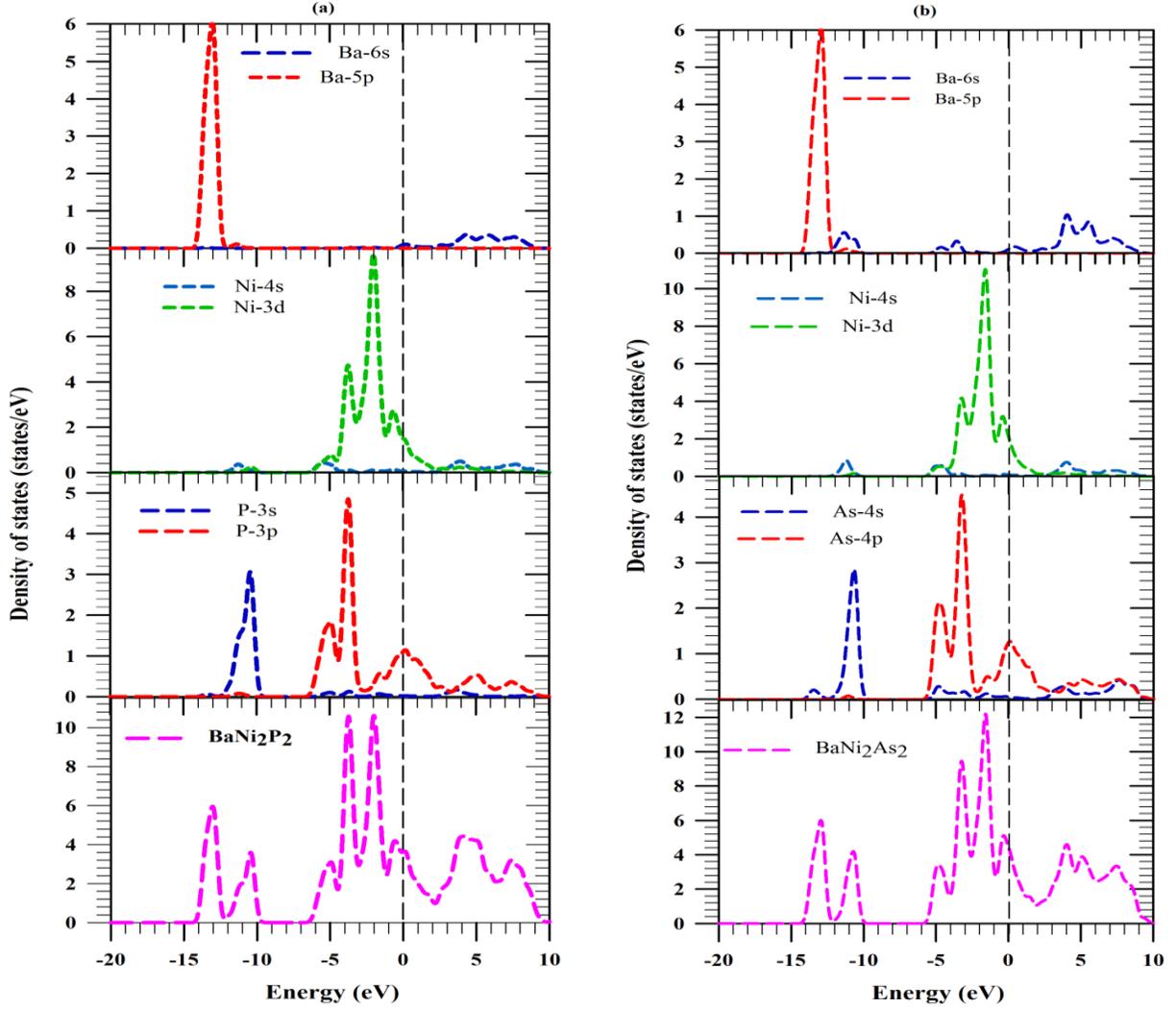

**Fig. 3.** Total and partial electronic density of states of (a) $BaNi_2P_2$ and (b) $BaNi_2As_2$.

The height valence bands for both the compounds are found to be in the energy range -12 eV to -15 eV come predominantly from Ba-5$p$ states. For both the phases a strong peak is found in the TDOS in the energy range -5 eV to -2 eV which is created from the strong hybridization of Ni-3$d$, P-3$p$ and As-4$p$ states indicating the covalent interaction among Ni-P and Ni-As bonds. At the Fermi level, the density of states mainly comes up from the Ni-3$d$ states. The conduction band is mostly originates from Ni-3$d$ and P-3$p$ states for $BaNi_2P_2$ and Ni-3$d$ and As-4$p$ states for $BaNi_2As_2$, where the major contribution comes from Ni-3$d$ state for both the superconductors. Our present calculations are well matched with the previous work [47].

In order to get a clear concept about the bonding nature of these two superconductors, the Mulliken atomic population [48-50] is calculated and given in Table 4. This study also provides the valuable information about the bond overlap and the transfer of charge among the different types of atoms in a material. From Table 4 it is understandable that, Ba atom contains positive charge while Ni and P atoms contain negative charges for BaNi$_2$P$_2$ superconductor representing the transfer of charge from Ba atom to Ni and P atoms. But in the case of BaNi$_2$As$_2$ superconductor, Ba and As atoms hold positive charge and only Ni atoms contain negative charge. Therefore charge transfer occurs from Ba and As atoms to Ni atom. Our calculations are well accord with the other similar types of theoretical investigation [51].

**Table 4.** Mulliken atomic populations of BaNi$_2$P$_2$ and BaNi$_2$As$_2$ superconductors.

| Compounds | Species | s | p | d | Total | Charge | Bond | Population | Lengths (Å) |
|---|---|---|---|---|---|---|---|---|---|
| BaNi$_2$P$_2$ | Ba | 1.67 | 6.04 | 1.05 | 8.76 | 1.24 | P-Ni | 0.59 | 2.249 |
| | Ni | 0.61 | 0.55 | 8.90 | 10.06 | -0.06 | Ni-Ni | -0.68 | 2.785 |
| | P | 1.82 | 3.74 | 0.00 | 5.56 | -0.56 | | | |
| BaNi$_2$As$_2$ | Ba | 2.66 | 6.04 | 1.07 | 9.78 | 0.22 | Ni-As | -0.68 | 2.358 |
| | Ni | 0.84 | 0.83 | 8.91 | 10.58 | -0.58 | Ni-Ni | -0.74 | 2.934 |
| | As | 0.76 | 3.77 | 0.00 | 4.53 | 0.47 | | | |

The nature of bonding between the different atoms is also confirmed by the study of total charge density of a compound. Here the total charge density of both the superconductors is investigated and depicted in Fig.4

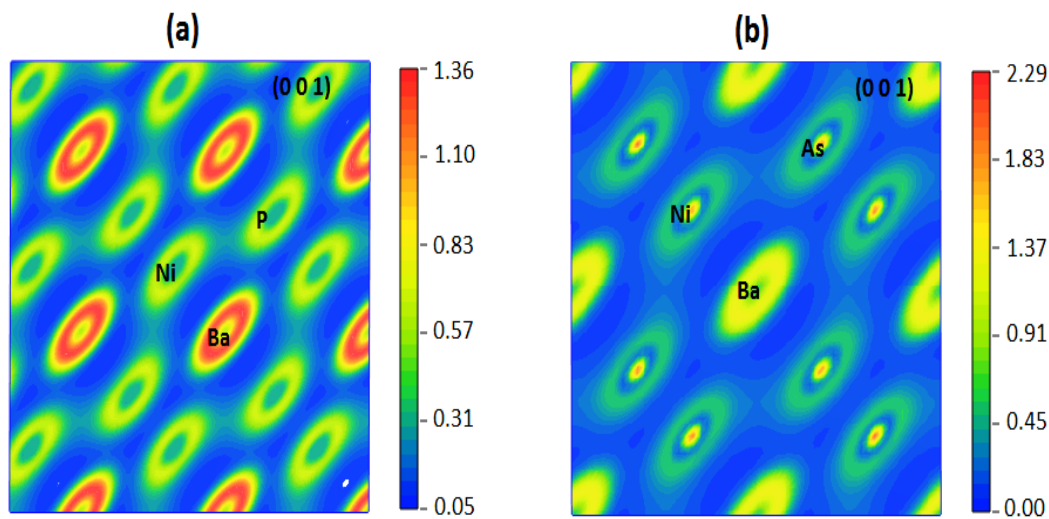

**Fig. 4.** The total charge density of (a) BaNi$_2$P$_2$ and (b) BaNi$_2$As$_2$ for (001) plane.

From Fig. 4 (a) it has been seen that a clear overlapping of charge distribution between Ni and P atoms is occurred indicating the existence of a covalent bond of Ni-P in BaNi$_2$P$_2$. Similarly for BaNi$_2$As$_2$ phase, a covalent feature of Ni-As bond is appeared. For both the compounds there is no overlapping of charge distribution among Ba atoms viewing the ionic character of Ba-Ba bond. The ionic nature is the consequence of the metallic character [52] showing the metallic feature of Ba-Ba bonds [51].

Overall, the comprehensive investigation of the density of states, Mulliken atomic population, and the total charge density of $BaNi_2P_2$ and $BaNi_2As_2$ superconductors point to the existence of all the ionic, covalent and metallic bonds in these compounds which is the common feature of $ThCr_2Si_2$-type compounds [50].

### 3.4 Optical properties

The investigation of the optical functions of solids is so imperative due to the fact that it helps to get a clear concept concerning the electronic configuration of materials. We have premeditated the optical functions of $BaNi_2T_2$ (T = P, As) with different photon energies through frequency-dependent dielectric function $\varepsilon(\omega) = \varepsilon_1(\omega) + i\varepsilon_2(\omega)$, which is intimately correlated to the electronic configurations. The imaginary section $\varepsilon_2(\omega)$ of this function is articulated as the momentum matrix constituents involving the filled and unfilled electronic states and can be easily calculated via [53]:

$$\varepsilon_2(\omega) = \frac{2e^2\pi}{\Omega\varepsilon_0} \sum_{k,v,c} \left|\langle \psi_k^c | \hat{u} \cdot \vec{r} | \psi_k^v \rangle\right|^2 \delta(E_k^c - E_k^v - E) \tag{12}$$

wherever $\omega$ refers to light frequency, $e$ indicates the electronic charge, $\hat{u}$ is the vector representing the polarization of the incident electric field, along with $\psi_k^c$ and $\psi_k^v$ are the conduction band and valence band wave functions at $k$, successively. The real part is calculated from the imaginary part via the Kramers-Kronig transforms. Besides this all other optical functions, for examples reflectivity, absorption coefficient, refractive index, conductivity, as well as loss function are derived by Eqs.(49)-(54) in Ref. [53].

Figs. 5 & 6 illustrate the calculated optical functions of $BaNi_2P_2$ and $BaNi_2As_2$ phases respectively with photon energies up to 40 eV for the polarization direction [100]. For all computations we have employed 0.5 eV Gaussain smearing since this spreads out the Fermi plane thus k-points will be more efficient at the Fermi surface.

Since both of the phases have no band gap as obvious from band structures, the photoconductivity begins with zero photon energy as depicted in Figs. 5(a) & 6 (a). From these figures we have seen that the conductivity spectra have a number of maxima and minima within the energy range 0-40 eV. For both of the phases we observed similar peak in the conductivity scheme. However, the photoconductivity and so the electrical conductivity of these two phases increases as a result of absorbing photons [54].

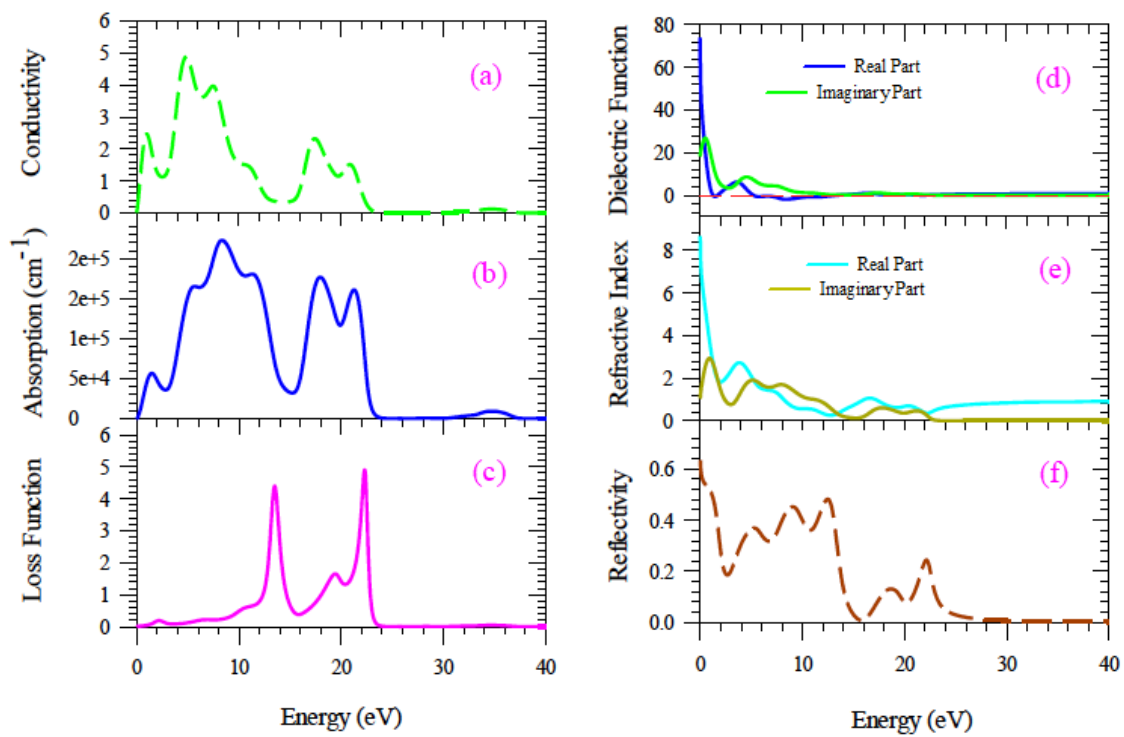

**Fig. 5:** The calculated optical functions of $BaNi_2P_2$: (a) conductivity, (b) absorption, (c) loss function, (d) dielectric function, (e) refractive index and (f) reflectivity

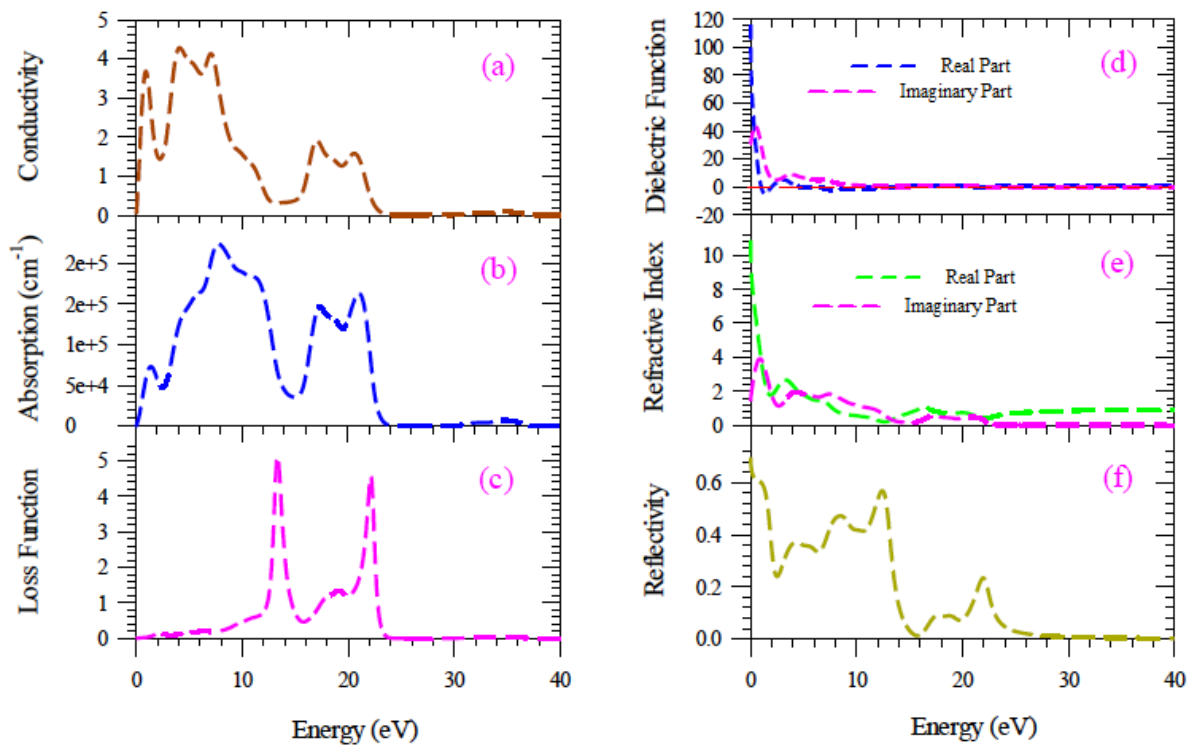

**Fig. 6.** The investigated optical functions of $BaNi_2As_2$: (a) conductivity, (b) absorption, (c) loss function, (d) dielectric function, (e) refractive index and (f) reflectivity.

As a function of photon energy the absorption spectra of $BaNi_2T_2$ (T = P, As) are shown in Figs. 5(b) & 6(b). It supplies the information about the optimum solar energy conversion efficiency and point out the penetration depth of light of precise energy into the material before being absorbed [55]. For both compounds the absorption spectra start at zero photon energy which reveals the metallic behaviors of these phases under study. Several peaks are viewed in the spectra but the height peak for $BaNi_2P_2$ is located at 4.63 eV and at 4.17 eV for $BaNi_2As_2$. Both superconducting phases exhibit quite good absorption coefficient in the energy range 4-10 eV.

The energy loss function is a key factor in the dielectric formalism utilized to clarify the optical spectra and the excitations created by fast charges in solid material. It is defined as the energy loss by fast electron traversing in the material [56]. The energy loss functions of these phases under study are depicted in Figs. 5(c) & 6(c). The frequency connected to the upper limit of the energy loss spectrum is specified by the bulk plasma frequency $\omega_p$ of the material, which emerges at $\varepsilon_2 < 1$ and $\varepsilon_1 = 0$ [57, 58]. From the energy loss spectra we have seen that the plasma frequency of $BaNi_2P_2$ and $BaNi_2As_2$ are come to 22.23 eV and 13.28 eV respectively, which reveals that these materials will be transparent if the incident photon has energy greater than the energy mentioned above.

An important optical function of solid material is the dielectric function which illustrates how a material reacts to the incident electromagnetic wave. As function of photon energy up to 40 eV the real and imaginary parts of dielectric functions of $BaNi_2P_2$ and $BaNi_2As_2$ are depicted in Figs. 5(d) and 6(d). The static values of $BaNi_2P_2$ and $BaNi_2As_2$ are 78 and 119 respectively. A metal with high dielectric constant is used to manufacture high value capacitors [59]. Hence in this case $BaNi_2As_2$ is more useful to manufacture this type of capacitors than $BaNi_2P_2$. From Figs. 5 and 6 (d), it is evident that the real part of the dielectric function tends to be unity in the high energy region and the imaginary part reaches nearly to zero [60]. It signifies that in this region these materials become almost transparent with small absorption.

The behavior of light or any other radiation through an optical medium is characterized by a dimensionless figure known as index of refraction. The refractive indices and the extinction coefficients of $BaNi_2P_2$ and $BaNi_2As_2$ are displayed in Figs. 5(e) and 6 (e) correspondingly. For these two phases the static refractive indices for the polarization vector [100] are 8.66 and 10.87 respectively. From these Figs. it is evident that the refractive indices of these phases are high in the infrared area and slowly reduced in the visible and ultraviolet regions.

Figs. 5(f) and 6 (f) represent the reflectivity spectra of $BaNi_2T_2$ as a function of incident photon energy. It has been evident that the reflectivity spectra for the phases $BaNi_2P_2$ and $BaNi_2As_2$ reveal nearly similar characteristics. For $BaNi_2P_2$ the reflectivity spectrum starts with a value of 0.63, at the beginning it decreases and then rises again to reach maximum value of 0.47 at 12.29 eV. The reflectivity spectrum of $BaNi_2As_2$ begins with a value of 0.696 decreases first and increases again to reach maximum value of 0.55 at 12.30 eV [61]. Hence both phases have almost alike reflectivity spectra, illustrate promise as excellent coating materials between 3.78 eV and 12.30 eV regions.

### 3.5 Thermodynamic properties

The thermodynamic properties of crystalline materials are very fundamental factors which explaine many solid-state phenomena directly. In this branch, the Debye temperature, thermal conductivity, melting temperature and Dulong-Petit limit of $BaNi_2P_2$ and $BaNi_2As_2$ have been calculated. The calculated values are listed in Table 6 and 7. The temperature that is connected to the highest frequency mode of vibration is called Debye temperature ($\theta_D$). The Debye temperature is directly connected to the thermal properties of material as like melting point of solid, specific heat of solid, thermal expansion etc. The Debye temperatures for superconductors $BaNi_2P_2$ and $BaNi_2As_2$ have been calculated using the following equation [62],

$$\theta_D = \frac{h}{k_B}\left(\frac{3N}{4\pi V}\right)^{\frac{1}{3}} \times v_m \qquad (13)$$

where, $h$ is the Planck constant, $k_B$ is the Boltzmann constant, $N$ is the number of atoms in the unit cell, V is the volume of the unit cell and $v_m$ is the average sound velocity that is given by,

$$v_m = \left[\frac{1}{3}\left(\frac{2}{v_t^3} + \frac{1}{v_l^3}\right)\right]^{-\frac{1}{3}} \qquad (14)$$

Where, $v_t$ is the transverse wave velocity and $v_l$ is the longitudinal wave velocity which are given by,

$$v_l = \left(\frac{3B + 4G}{3\rho}\right)^{\frac{1}{2}} \qquad (15)$$

$$v_t = \left(\frac{G}{\rho}\right)^{\frac{1}{2}} \qquad (16)$$

The calculated values of $\rho$, $v_t$, $v_l$ and $v_m$ of $BaNi_2P_2$ and $BaNi_2As_2$ are shown in Table 6. Using these values in Eq. (13) the obtained Debye temperatures ($\theta_D$) of both the superconductors $BaNi_2P_2$ and $BaNi_2As_2$ are 323.70 K and 272.94 K respectively.

**Table** 6: The calculated density $\rho$ (in gm/cm$^3$), transverse ($v_t$), longitudinal ($v_l$), and average sound velocity $v_m$ (m/s) and Debye temperature $\Theta_D$ (K) of BaNi$_2$P$_2$ and BaNi$_2$As$_2$ superconductors.

| Compounds | $\rho$ | $v_t$ | $v_l$ | $v_m$ | $\Theta_D$ | Remarks |
|---|---|---|---|---|---|---|
| BaNi$_2$P$_2$ | 5.68 | 2513.23 | 4761.23 | 2876.93 | 323.70 | This study |
|  | - | - | - | - | 290.00 | Exp. [67] |
| BaNi$_2$As$_2$ | 6.87 | 2204.07 | 4297.99 | 2468.74 | 272.94 | This study |
|  | - | - | - | - | 250.00 | Exp. [68] |

The temperature at which solid turns into liquid at atmospheric pressure is called melting point. The melting temperature of tetragonal crystal can be obtained by using Eq. 17 [63],

$$T_m = 354 + \frac{4.5\ (2C_{11} + C_{33})}{3} \tag{17}$$

The calculated melting temperatures of BaNi$_2$T$_2$ (T= P, As) are listed in Table 7. It is clear that the calculated melting temperature of BaNi$_2$P$_2$ is greater than BaNi$_2$As$_2$ superconductor which reveals that the stability of BaNi$_2$P$_2$ superconductor is more than the BaNi$_2$As$_2$ superconductor.

Thermal conductivity is the most important properties of a solid which is responsible for conducting heat energy. In this work, the minimum thermal conductivity can be obtained by using the following expression [64, 65],

$$K_{min} = K_B v_m \left(\frac{M}{n\rho N_A}\right)^{-\frac{2}{3}} \tag{18}$$

where, $K_B$ indicates the Boltzmann constant, $v_m$ denotes the average sound velocity, M represents the molecular mass, $n$ is defined as the number of atoms per molecule and $N_A$ is the Avogadro's numbers. The calculated values of the minimum thermal conductivity $K_{min}$ of BaNi$_2$P$_2$ and BaNi$_2$As$_2$ superconductors are shown in Table 7. Both superconductors possess relatively low thermal conductivity at ambient condition.

**Table 7**: The calculated melting temperature, $T_m$ (K), the minimum thermal conductivity, $K_{min}$ (in $Wm^{-1}K^{-1}$) and the Dulong-petit limit (J/mole.K) of $BaNi_2P_2$ and $BaNi_2As_2$ superconductors.

| Compounds | $T_m \pm 300$ | $K_{min}$ | Dulong-petit limit |
|---|---|---|---|
| $BaNi_2P_2$ | 1003.60 | 0.56 | |
| $BaNi_2As_2$ | | | 124.67 |
| | 981.99 | 0.46 | |

At high temperature and at a constant volume, the anharmonic effect of the specific heat capacity $C_v$ of a solid element is subdued and close to a limit at high temperature is known as Dulong-petit limit [66]. The Dulong-petit limit can be evaluated by the following relation [66],

$$Dulong - petit\ limit = 3nN_AK_B \qquad (19)$$

where, $N_A$ and $K_B$ represent the Avogadro's numbers and Boltzmann constant respectively. The evaluated value of Dulong-petit limit of both superconductors is shown in Table 7.

## 4. Conclusions

In the present research work, we have performed an *ab-initio* simulation depended on the density functional theory to investigate the different physical properties including structural, elastic, electronic, optical and thermodynamic properties of two nickel based superconductors $BaNi_2P_2$ and $BaNi_2As_2$. The optimized lattice constants reveal good accordance with the experimental results. The investigation of the mechanical properties reveals that both the compounds are mechanically stable and show anisotropic activities. The calculated B/G values ensure that both the compounds are ductile in nature and the Poisson's ratio demonstrates the existence of central force in $BaNi_2P_2$ and $BaNi_2As_2$. The calculations of band structures and density of states indicate that both the compounds have metallic behaviors. The study of the Mulliken atomic populations and total charge density ensure the existence of ionic, covalent and metallic bonds in both the compounds. The investigation of the optical functions reveals that both the compounds have high refractive indices in the infrared region and gradually decreased in the visible and ultraviolet regions. For both these phases the conductivity spectrum and the absorption coefficient are started from zero energy which indicates the metallic features of these compounds. These characteristics are also observed from the band structure and DOS calculations. The study of the dielectric constant confirms that $BaNi_2As_2$ is good dielectric material than $BaNi_2P_2$. The large reflectivity in the low energy site indicates that these compounds may be used as coating material to avoid solar heating. The calculated Debye temperatures of BaNi$_2$P$_2$ and BaNi$_2$As$_2$ superconductors are 323.70 K and 272.94 K respectively which shows good agreement with the experimental results. Finally the calculated thermal conductivities of BaNi$_2$P$_2$ and BaNi$_2$As$_2$ compounds are 0.56 and 0.46 respectively.

**References**

1. Hirai, Daigorou, et al., Superconductivity in Layered Pnictides BaRh$_2$P$_2$ and BaIr$_2$P$_2$, Journal of the Physical Society of Japan 78.2 (2009), 023706-023706.
2. Bednorz, J. George, and K. Alex Müller, Possible highTc superconductivity in the Ba− La− Cu− O system, Zeitschrift für Physik B Condensed Matter 64.2 (1986), 189-193.
3. Kamihara Y., Watanabe T., Hirano M. and Hosono H., *J. Am. Chem. Soc.*, 130 (2008) 3296.
4. Ren Z. A., Che G. C., Dong X. L., Yang J., Lu W., Yi W., Shen X. L., Li Z. C., Sun L. L., Zhou F. and Zhao Z. X., *EPL*, 83 (2008) 17002.
5. Guo, Qi, et al. "Superconductivity at 3.85 K in BaPd$_2$As$_2$ with the ThCr$_2$Si$_2$-type structure." EPL (Europhysics Letters) 113.1 (2016): 17002.
6. Guo J. G., Jin S. F., Wang G., Wang S., Zhu K., Zhou T., He M. and Chen X. L., Phys. Rev. B, 82 (2010) 180520(R).
7. Johnston D. C., Adv. Phys., 59 (2010) 803.
8. Sefat A. S., Jin R. Y., McGuire M. A., Sales B. C., Singh D. J. and Mandrus D., Phys. Rev. Lett., 101 (2008) 117004.
9. Stewart G. R., Rev. Mod. Phys., 83 (2011) 1589.
10. M. Rotter, M. Tegel, D. Johrendt, Superconductivity at 38 K in the iron arsenide (Ba$_{1-x}$K$_x$)Fe$_2$As$_2$, Phys. Rev. Lett. 101 (2008) 107006.
11. K. Sasmal, B. Lv, B. Lorenz, et al., Superconducting Fe-based compounds (A$_{1-x}$ Sr$_x$)Fe$_2$As$_2$ with A = K and Cs with transition temperatures up to 37 K, Phys. Rev. Lett. 101 (2008) 107007.
12. M.S. Torikachvili, S.L. Bud'ko, N. Ni, et al., Pressure induced superconductivity in I$_2$As$_2$, Phys. Rev. Lett. 101 (2008) 057006
13. D.C. Johnston, The puzzle of high temperature superconductivity in layered iron pnictides and chalcogenides, Adv. Phys. 59 (2010) 803-1061.
14. G.R. Stewart, Superconductivity in iron compounds, Rev. Mod. Phys. 83 (2011) 1589-1652.
15. Keimes, V., et al. "Zur Polymorphie von SrNi$_2$P$_2$ sowie zur Kristallstruktur von BaNi$_2$P$_2$." Zeitschrift für anorganische und allgemeine Chemie 623.11 (1997): 1699-1704.

16. Mine, Takashi, et al. "Nickel-based phosphide superconductor with infinite-layer structure, $BaNi_2P_2$." Solid State Communications 147.3 (2008): 111-113.

17. Shein, I. R., and A. L. Ivanovskii. "Electronic and structural properties of low-temperature superconductors and ternary pnictides $ANi_2Pn_2$ (A = Sr, Ba and P n = P, As)." Physical Review B 79.5 (2009): 054510.

18. Ronning, F., et al. "The first order phase transition and superconductivity in $BaNi_2As_2$ single crystals." Journal of Physics: Condensed Matter 20.34 (2008): 342203.

19. Sefat, Athena S., et al. "$BaT_2As_2$ single crystals (T= Fe, Co, Ni) and superconductivity upon Co-doping." Physica C: Superconductivity 469.9 (2009): 350-354.

20. Kurita, N., et al. "Fully gapped superconductivity in Ni-pnictide superconductors $BaNi_2As_2$ and $SrNi_2P_2$." Journal of Physics: Conference Series. Vol. 273. No. 1. IOP Publishing, 2011.

21. Subedi, Alaska, and David J. Singh. "Density functional study of $BaNi_2As_2$: Electronic structure, phonons, and electron-phonon superconductivity." Physical Review B 78.13 (2008): 132511.

22. Rahaman, Md Zahidur, and Md Atikur Rahman. "$ThCr_2Si_2$-type Ru-based superconductors $LaRu_2M_2$ (M = P and As): An ab-initio investigation." Journal of Alloys and Compounds 695 (2017): 2827-2834.

23. Rahaman, Md Zahidur, and Md Atikur Rahman. "Novel 122-type Ir-based superconductors $BaIr_2Mi_2$ (Mi = P and As): A density functional study." Journal of Alloys and Compounds 711 (2017): 327-334.

24. W. Kohn, L.J. Sham, Phys. Rev. A 140 (1965) 1133.

25. S.J. Clark, M.D. Segall, C.J. Pickard, P.J. Hasnip, M.J. Probert, K. Refson, M.C. Payne, Z. Kristallogr. 220 (2005) 567.

26. J.P. Perdew, K. Burke, M. Ernzerhof, Phys. Rev. Lett. 77 (1996) 3856.

27. K.J. Monkhorst, J.D. Pack, Phys. Rev. B 13 (1976) 5188.

28. T.H. Fischer, J.A. lmlof, J. Phys. Chem. 96 (1992) 768.

29. R. Hill, The elastic behavior of a crystalline aggregate, Proc. Phys. Soc. A 65 (1952) 349e355.

30. S. Rozsa, H.U. Schuster, Z. Naturforsch, B: Chem. Sci. 36 (1981) 1668.

31. Czybulka, A., M. Noack, and H-U. Schuster. "Neue ternäre Phosphide und Arsenide des Caesiums mit Elementen der 8. Nebengruppe." Zeitschrift für anorganische und allgemeine Chemie 609.3 (1992), 122-126.

32. Keimes, V., et al. "Zur Polymorphie von $SrNi_2P_2$ sowie zur Kristallstruktur von $BaNi_2P_2$, Zeitschrift für anorganische und allgemeine Chemie 623.11 (1997), 1699-1704.

33. Pfisterer, Martin, and Günter Nagorsen. "Zur Struktur ternärer Übergangsmetallarsenide, On the Structure of Ternary Arsenides." Zeitschrift für Naturforschung B 35.6 (1980), 703-704.

34. Pfisterer, Martin, and Günter Nagorsen. "Bindungsverhältnisse und magnetische Eigenschaften ternärer Arsenide ET$_2$As$_2$, Bonding and Magnetic Properties in Ternary Arsenides ET$_2$As$_2$, Zeitschrift für Naturforschung B 38.7 (1983), 811-814.
35. Y.D. Zhu, et al., First-principles investigation of structural, mechanical and electronic properties for CueTi intermetallics, Comput. Mater. Sci. 123 (2016) 70-78.
36. Md. Zahidur Rahaman, Md. Atikur Rahman, Computational Condensed Matter 8 (2016) 7-13.
37. J.F. Nye, Propriété´s Physiques des Maté´riaux, Dunod, Paris, 1961.
38. Piskunov, S., et al., Bulk properties and electronic structure of SrTiO$_3$, BaTiO$_3$, PbTiO$_3$ perovskites: an ab initio HF/DFT study, Computational Materials Science 29.2 (2004): 165-178.
39. Hill, Richard, The elastic behaviour of a crystalline aggregate, Proceedings of the Physical Society, Section A 65.5 (1952), 349.
40. Ranganathan, Shivakumar I., and Martin Ostoja-Starzewski, Universal elastic anisotropy index, Physical Review Letters101.5 (2008), 055504.
41. Chen, Xing-Qiu, et al., Modeling hardness of polycrystalline materials and bulk metallic glasses, Intermetallics 19.9 (2011), 1275-1281.
42. Ravindran, P., et al., Density functional theory for calculation of elastic properties of orthorhombic crystals: application to TiSi$_2$, Journal of Applied Physics 84.9 (1998), 4891-4904.
43. Pugh, S. F., Relations between the elastic moduli and the plastic properties of polycrystalline pure metals, The London, Edinburgh, and Dublin Philosophical Magazine and Journal of Science 45.367 (1954), 823-843.
44. Cao, Yong, et al., First-principles studies of the structural, elastic, electronic and thermal properties of Ni$_3$Si, Computational Materials Science 69 (2013), 40-45.
45. WC Hu, Y Liu, DJ Li, XQ Zeng, CS Xu, First-principles study of structural and electronic properties of C14-type Laves phase Al$_2$Zr and Al$_2$Hf , Computational Materials Science, (2014).
46. M.Z. Rahaman, M.A. Rahman, Novel Laves phase superconductor NbBe$_2$: a theoretical investigation, Comput. Condens. Matter 8 (2016) 7–13.
47. I. R. Shein and A. L. Ivanovskii, Physical review B 79, 054510(2009).
48. R.S. Mulliken, J. Chem. Phys. 23 (1955) 1833.
49. D. Sanchez-Portal, E. Artacho, J.M. Soler, Solid State Commun. 95 (1995) 685.
50. M.D. Segall, R. Shah, C.J. Pickard, M.C. Payne, Phys. Rev. B 54 (1996) 16317.
51. Md. Zahidur Rahaman, Md. Atikur Rahman, Journal of Alloys and Compounds 695 (2017) 2827-2834.
52. R.P. Singh, J. Magnesium Alloys 2 (2014) 349-356.

53. Materials Studio CASTEP Manual © Accelrys, 2010. http://www.tcm.phy.cam.ac.uk/castep/documentation/WebHelp/CASTEP.html.
54. Sun, J., Zhou, X. F., Fan, Y. F., Chen, J. and Wang, H. T. [2006] "First-principles study of electronic structure and optical properties of heterodiamond $BC_2N$," Phys. Rev. B 73, 045108–045110.
55. Md. Lokman Ali, Md. Zahidur Rahaman and Md. Atikur Rahman, The structural, elastic and optical properties of ScM (M = Rh, Cu, Ag, Hg) intermetallic compounds under pressure by ab initio simulations, Vol. 5, No. 4 (2016) 1650024.
56. M. Xu, S.Y. Wang, G. Yin, J. Li, Y. X. Zheng, L.Y. Chen, and Y. Jia Appl. Phys. Lett. 89 (2006) 151908.
57. R. Saniz, L.H. Ye, T. Shishidou, and A. Freeman, Phys. Rev. B 74 (2006) 014209.
58. de J.S. Almeida and R. Ahuja Phys. Rev. B 73 (2006) 165102.
59. Md. Afjalur Rahman, Md. Zahidur Rahaman, and Md. Atikur Rahman, Int. J. Mod. Phys. B DOI: http://dx.doi.org/10.1142/ S021797921650199X.
60. M. Roknuzzaman, M.A. Hadi, Physical properties of predicted $Ti_2CdN$ versus existing $Ti_2CdC$ MAX phase: An ab initio study, Computational Materials Science 113 (2016) 148–153.
61. Md. Atikur Rahman, Md. Zahidur Rahaman, et al.: First principles investigation of structural, elastic, electronic and optical properties of $HgGeB_2$ (B=P, As) chalcopyrite semiconductors, Computational Condensed Matter 9(2016) 19-26, http://dx.doi.org/10.1016/j.cocom.2016.09.001.
62. Aydin, Sezgin, and Mehmet Simsek, First-principles calculations of $MnB_2$, $TcB_2$, and $ReB_2$ within the $ReB_2$-type structure, Physical Review B 80.13 (2009), 134107.
63. Alouani, M., R. C. Albers, and Michael Methfessel, Calculated elastic constants and structural properties of Mo and $MoSi_2$, Physical Review B43.8 (1991), 6500.
64. Y. Shen, D.R. Clarke and P.P.A. Fuierer, Anisotropic thermal conductivity of the aurivillus phase, bismuth titanate ($Bi_4Ti_3O_{12}$). A natural nanostructured superlattice: Appl. Phys. Lett. 93 (2008) pp. 102907–3.
65. Clarke, David R. Materials selection guidelines for low thermal conductivity thermal barrier coatings, Surface and Coatings Technology 163 (2003), 67-74.
66. Mao, Xiao-Chun, et al. "Theoretical Investigation of the Structural, Elastic, and Thermodynamic Properties of $MgAl_2O_4$ Spinel under High Pressure." Journal of the Physical Society of Japan 85.11 (2016): 114605.
67. D Hirai, F V Rohr and R J Cava Phys. Rev. B 86 100505 (2012)
68. A.S. Sefat, R. Jin, M.A. McGuire, B.C. Sales, D. Mandrus, Structure and Anisotropic Properties of $BaFe_{2-x}Ni_xAs_2$(x = 0, 1, 2) Single Crystals, 1-17.